\documentclass[prb,twocolumn,superscriptaddress,showpacs,floatfix]{revtex4}
\usepackage{amssymb,amsmath}
\usepackage{braket}
\usepackage{graphicx}

\newcommand{\lkv}{\left[}
\newcommand{\rkv}{\right]}
\newcommand{\lfi}{\left\{}
\newcommand{\rfi}{\right\}}

\newcommand{\bsl}[1]{\begin{slide}{#1}}
\newcommand{\esl}{\end{slide}}
\newcommand{\be}{\begin{equation}}
\newcommand{\ee}{\end{equation}}
\newcommand{\ben}{\begin{enumerate}}
\newcommand{\een}{\end{enumerate}}
\newcommand{\bit}{\begin{itemize}}
\newcommand{\eit}{\end{itemize}}
\newcommand{\been}{\begin{displaymath}}
\newcommand{\eeen}{\end{displaymath}}
\newcommand{\ba}{\left[\begin{array}}
\newcommand{\ea}{\end{array}\right]}
\newcommand{\bac}{\begin{array}}
\newcommand{\eac}{\end{array}}
\newcommand{\bc}{\begin{center}}
\newcommand{\ec}{\end{center}}
\newcommand{\bea}{\begin{eqnarray}}
\newcommand{\eea}{\end{eqnarray}}
\newcommand{\bean}{\begin{eqnarray*}}
\newcommand{\eean}{\end{eqnarray*}}
\newcommand{\bqu}{\begin{quote}\begin{it}}
\newcommand{\equ}{\end{it}\end{quote}}

\newcommand{\vk}{\mathbf{k}}

\newcommand{\vvr}{\mathbf{r}}

\newcommand{\vvR}{\mathbf{R}}

\DeclareMathOperator{\Tr}{Tr}

\begin{document}
\bibliographystyle{apsrev}
\title{Correlation effects in total energy of transition metals and related properties}
\author{I. Di Marco}
\email{dimarco@science.ru.nl}
\affiliation{Institute for Molecules and Materials, Radboud University of Nijmegen, NL-6525 ED Nijmegen, The Netherlands}
\author{J. Min\'{a}r}
\affiliation{Department Chemie und Biochemie, Physikalische Chemie, Ludwig-Maximilians Universit\"{a}t M\"{u}nchen, D-81377 M\"{u}nchen, Germany}
\author{S. Chadov}
\affiliation{Department Chemie und Biochemie, Physikalische Chemie, Ludwig-Maximilians Universit\"{a}t M\"{u}nchen, D-81377 M\"{u}nchen, Germany}
\affiliation{Institut f\"{u}r Anorganische und Analytische Chemie, Johannes-Gutenberg Universit\"{a}t  Mainz, 55128 Mainz, Germany}
\author{M. I. Katsnelson}
\affiliation{Institute for Molecules and Materials, Radboud University of Nijmegen, NL-6525 ED Nijmegen, The Netherlands}
\author{H. Ebert}
\affiliation{Department Chemie und Biochemie, Physikalische Chemie, Ludwig-Maximilians Universit\"{a}t M\"{u}nchen, D-81377 M\"{u}nchen, Germany}
\author{A. I. Lichtenstein}
\affiliation{Institute of Theoretical Physics, University of Hamburg, 20355 Hamburg, Germany}

\date{\today}

\begin{abstract}
We present an accurate implementation of total energy calculations
into the local density approximation plus dynamical mean-field theory
(LDA+DMFT) method. The electronic structure problem is solved
through the full potential linear Muffin-Tin Orbital (FP-LMTO) and
Korringa-Kohn-Rostoker (FP-KKR) methods with a perturbative solver for the effective impurity
suitable for moderately correlated systems. We have tested the
method in detail for the case of Ni and investigated the
sensitivity of the results to the computational scheme and to the complete
self-consistency. It is demonstrated that the LDA+DMFT method can
resolve a long-standing controversy between the LDA/GGA density
functional approach and experiment for equilibrium lattice constant and
bulk modulus of Mn.
\end{abstract}
\maketitle

\section{Introduction}
The state-of-the-art technique for calculating the electronic
structure of materials is density functional
theory\cite{jones89rmp61:689,dreizler-gross-book-1990} in its local
density approximation (LDA). However,
despite numerous impressive successes, it faces serious
difficulties for strongly correlated systems such as Mott
insulators, heavy fermion systems, high-temperature
superconductors, itinerant electron magnets, and many others. Some
of these difficulties were overcame by merging LDA-based first-principles
electronic structure calculations with the dynamical mean-field
theory (the LDA+DMFT
approach\cite{anisimov97jpcm9:7359,lichtenstein98prb57:6884}; for
review see Refs.
\onlinecite{kotliar04pt57:53,kotliar06rmp78:865,held07ap56:829,katsnelson08rmp80:315}).
Most of the works done by this method deal with spectral
properties of strongly correlated systems. At the same time,
correlation effects are sometimes of crucial importance to
describe also cohesive energy, equilibrium lattice constant and
bulk modulus, as demonstrated for the cases of
plutonium\cite{savrasov01nature410:793, savrasov04prb69:245101}
and cerium\cite{held01prl87:276404,amadon06prl96:066402}. After
these first attempts, the need of more systematic implementations
and investigation of the numerical problems related to total
energy evaluation in the LDA+DMFT scheme arose. Recently
Pourovskii et al\cite{pourovskii07prb76:235101} have presented an
interesting comparison between the correlation effects in the
basic DMFT cycle (convergence in the local self-energy) and in the
fully self-consistent DMFT cycle (convergence in the local
self-energy and in the electron density). Two test-cases have been
studied with this new implementation: the $\gamma$ phase of
metallic cerium and the Mott insulator Ce$_2$O$_3$. Both of them
are close-packed $f$-electron systems and they can be studied
through the atomic sphere approximation within the linear
muffin-tin orbital method (ASA-LMTO) and through the Hubbard-I
solver\cite{lichtenstein98prb57:6884}.

Up to now all the LDA+DMFT studies of the ground state properties
of strongly correlated systems concerned materials with rather
localized $f$-electrons. Here, we are interested in
materials where the correlation effects are less dramatic and
where the failures of the density-functional theory pertain only
some specific properties. The late transition metals Mn, Fe, Co, and
Ni are correlated systems, and the LDA+DMFT approach was
successfully applied to describe their spectral
properties\cite{katsnelson99jpcm11:1037,katsnelson02epjb30:9,
katsnelson00prb61:8906,lichtenstein01prl87:067205,biermann04zhetfl80:714,
minar05prb72:045125, minar05prl95:166401, braun06prl97:227601,
grechnev07prb76:035107} as well as their magnetic
properties\cite{lichtenstein01prl87:067205, chadov08epl82:37001}.
In particular, the DMFT was implemented into full-potential
Korringa-Kohn-Rostoker method (FP-KKR)\cite{minar05prb72:045125}
and full-potential linear muffin-tin orbital method
(FP-LMTO)\cite{grechnev07prb76:035107} to allow corresponding studies.

In the present paper we extend the previous implementations to calculate
the total energy of the electronic system whitin the LDA+DMFT scheme. While
there exists already another LDA+DMFT code based on FP-LMTO and able to
calculate ground state properties of strongly correlated
materials\cite{savrasov04prb69:245101}, we must emphasize that it is
the first time that this happens for an LDA+DMFT code based on FP-KKR.
This is particularly appealing to analyze disordered alloys systems,
for which FP-KKR in combination with coherent potential
approximation (CPA) alloy-theory is well-known
to be particularly efficient and reliable. 

Here we use these implementations to study the total energy and related
properties of 3$d$ transition metals. First, we present
computational results for Ni which plays the role of
``drosophila fly'' for the LDA+DMFT method and where the most
detailed comparison of the theory with experiment was
done\cite{minar05prl95:166401, braun06prl97:227601}. After
calculations of photoemission, optical and magnetooptical spectra,
magnetization, magnetic susceptibility, orbital
magnetic moments, bulk and surface spectral densities (see
previous works cited above), the present calculation of 
cohesive energy, equilibrium lattice constant and bulk modulus
completes its basic physical description within the LDA+DMFT
approach.

Comparing the results of the full-potential KKR and LMTO
calculations we address the question about sensitivity of the
LDA+DMFT description to the band structure method used. This is
nontrivial since different methods use different basis sets which
are truncated in any real calculations. We have found that
actually the results are very close which support its reliability.
While correlation effects in ground state properties of Ni
are quite small, they are accurately described within our scheme
which confirms the usefulness of the LDA+DMFT for not only strongly
correlated but also for moderately correlated systems. We have
checked also the importance of the full charge self-consistency and
found that in the case if Ni these effects are not very
essential.

Then, we consider the case of Mn where, among all transition
metals, the largest discrepancy between the LDA or GGA predictions
for the lattice constant and bulk modulus and the experimental
data takes place\cite{moruzzi93prb48:7665,eder00prb61:11492,
hafner05prb72:144420} which is considered to be an indication of
the strongest correlation effects among 3$d$
metals\cite{zein95prb52:11813,biermann04zhetfl80:714}. We show
that the LDA+DMFT method does allow us to solve this problem and
to describe in a very satisfactory way the energetics of Mn.

\section{Formulation of the problem}
All the standard approaches for calculating the electronic structure
of strongly correlated materials are based on the
choice of a set of orbitals that are described not accurately
enough in the standard DFT-LDA method which is supposed to be
improved. We call them ``correlated orbitals'' and indicate with
$\ket{\vvR, \xi}$, where $\vvR$ is the vector specifying the
Bravais lattice site and the $\xi$ is an index that enumerates the
orbitals within the unit cell of the crystal.  The choice of $\lfi
\ket{\vvR, \xi} \rfi$ is dictated by physical motivations for the
problem under consideration and always implies some degree of
arbitrariness (see the discussion below). Usually the correlated
orbitals are derived from $d$ or $f$ atomic states and the index
$\xi$ stands for the atomic quantum numbers $l$, $m$, $\sigma$.
Natural choices can be Linear Muffin-Tin
Orbitals\cite{savrasov04prb69:245101} or Wannier
functions\cite{anisimov05prb71:125119,lechermann06prb74:125120}.
Apart from the atomic states, hybridized orbitals can also be
chosen depending on the problem. For example in the transition
metal oxides the crystal field splits the LDA bands in two
distinct groups, well separated in energy and suitable to be
determined through downfolding of the original problem via
the NMTO
approach\cite{andersen00prb62:16219}.

After having decided the set $\lfi \ket{\vvR, \xi} \rfi$, we
correct the standard DFT-LDA Hamiltonian with an additional
Hubbard interaction term\cite{kotliar06rmp78:865} that explicitly
describes the local Coulomb repulsion $U$ for the orbitals in the
set:
\be
\label{eq:hlpu} {H} = {H}_{LDA} + \frac 12 \sum_{\vvR}
\sum_{\xi_1,\xi_2,\xi_3,\xi_4} U_{\xi_1,\xi_2,\xi_3,\xi_4}
c^{\dagger}_{\vvR,\xi_1} c^{\dagger}_{\vvR,\xi_2} c_{\vvR,\xi_4}
c_{\vvR,\xi_3}.
\ee

This is the so-called LDA+U Hamiltonian and an important remark
has to be made concerning the meaning of the matrix elements
$U_{\xi_1,\xi_2,\xi_3,\xi_4}$. We should not think of them as
generic matrix elements of the bare Coulomb repulsion, but more as
the matrix elements of an effective interaction introduced to give
the correct description of the low energy excitations (to describe
broader energy scales the $U$ term should be, in general, energy
dependent\cite{aryasetiawan04prb70:195104}). In these terms we
have to consider the LDA+U Hamiltonian as being derived
from completely  {\it ab initio} density functional approach.
While in principle this is possible through many methods, e.g.
constrained density functional
theory\cite{gunnarsson90prb41:514,anisimov91prb43:7570},
extraction from GW results\cite{aryasetiawan04prb70:195104}, it is
a common practice to evaluate the matrix elements
$U_{\xi_1,\xi_2,\xi_3,\xi_4}$ using semi-empirical
procedures\cite{oles84prb29:314,katsnelson99jpcm11:1037}. This may
seem to be inadequate, since the strength of the effective Coulomb
interaction should depend on the set of correlated orbitals, being
strictly connected to a mapping of the original electronic
Hamiltonian into the Eq. (\ref{eq:hlpu}); however if the orbitals
$\lfi \ket{\vvR, \xi} \rfi$ are chosen appropriately, the results are
quite stable with respect to this ambiguity, as it was first
noticed for the LDA+U method\cite{anisimov97jpcm9:767}.

If the correlated orbitals are atomic-like ones (with the quantum
numbers $l,m,\sigma$), we can express\cite{anisimov97jpcm9:767} the
Coulomb parameters in terms of Slater integrals $F^n$ like
 \be
 \label{eq:couslater}
  U_{\xi_1,\xi_2,\xi_3,\xi_4} =
  \sum_{n=0}^{2l}{a_n(\xi_1,\xi_3,\xi_2,\xi_4)F^n},
 \ee
with the coefficients $a_n$ defined as
 \be
 \label{eq:slatercoeff}
 a_n(\xi_1,\xi_3,\xi_2,\xi_4) = \frac{4\pi}{2n+1} \sum_{q=-n}^{+n}{ \braket{\xi_1 | Y_{nq} |\xi_3}
 \braket{\xi_2 | Y_{nq}^* |\xi_4} },
 \ee
where the terms $\braket{\xi_1 | Y_{nq} |\xi_3}$ and
$\braket{\xi_2| Y_{nq}^* |\xi_4}$ are integrals over products of
three spherical harmonics. In the rest of the paper we will be
interested in $3d$ electrons, therefore we can limit our discussion to
these. For $d$ electrons there are only three Slater parameters
($F^0$, $F^2$ and $F^4$) and they can be easily linked to the
Coulomb parameter $U$ and the Stoner parameter $J$
as\cite{anisimov97jpcm9:767} \be U=F^0, \quad
J=\frac{F^2+F^4}{14}. \ee The ratio $F^4/F^2$ is assumed to
correspond to the atomic value and for $3d$ electrons it is
approximately equal to $0.625$. In the rest of the paper we will
use the two real values $U$ and $J$ to specify the Coulomb matrix
elements.

\section{Dynamical Mean-Field Theory}
The LDA+U Hamiltonian defines an ``effective'' Hubbard model, and
its solution represents a complicated many-body problem. The
strategy of the spectral density functional
theory\cite{kotliar06rmp78:865} is the same of DFT or
Baym-Kadanoff theory (or more generally of every Weiss mean-field
theory): we specify a main observable quantity and we map the
original system into a system with less degrees of freedom under
the condition of conserving the expectation value of the main
observable. In DFT and Baym-Kadanoff theory the main observables
are respectively the total electron density $\rho(\vvr)$ and the
one electron Green's function $\hat{G}(z)$, namely \be
\label{eq:gf}
 \hat{G} (z) = \lkv  (z-\mu) \hat{\mathbf{1}} -  \hat{h}_{LDA} -  \hat{\Sigma}(z) \rkv^{-1},
\ee where $z$ is the energy in the complex plane, $\mu$ is the
chemical potential, $h_{LDA}$ plays the role of the unperturbed
Hamiltonian ("hopping"), and $\hat{\Sigma}(z)$ is the self-energy
operator, which in many-body theory reproduces the effects of the
interactions. In spectral density functional theory the main
observable is the local Green's function at the site $\vvR$: \be
\label{eq:gloc}
 \hat{G}_{\vvR}(z) =  \hat{P}_{\vvR}  \hat{G}(z)  \hat{P}_{\vvR},
\ee
where
\be
 \hat{P}_{\vvR} = \sum_{\xi} \ket{\vvR,\xi}\bra{\vvR,\xi}
\ee
is the projection operator to the correlated subspace belonging to site $\vvR$.

As in density functional theory, where we make approximations as LDA
or GGA, in the framework of the spectral density functional theory
the corresponding approximation is the dynamical mean-field theory. In the
DMFT the self-energy is considered to be purely local. In terms of
matrix elements on the correlated orbitals at the two sites
$\vvR_1$ and $\vvR_2$, this means that \be \label{eq:dmftsigma}
\braket{\vvR_1,\xi_1| \hat{\Sigma}(z) | \vvR_2,\xi_2}  =
\delta_{\vvR_1,\vvR_2} \braket{\xi_1| \hat{\Sigma}_{\vvR_1}(z)
|\xi_2} . \ee The assumption of a purely local self-energy
$\hat{\Sigma}_{\vvR}(z)$ allows us to focus only on the single
lattice site $\vvR$. Consequently we can replace all the other
sites of the lattice with a self-consistent electronic bath (or
``dynamical mean-field'') $\hat{\mathcal{G}}_0^{-1}(\vvR,z)$, whose
role is analogous to the Weiss mean-field used in statistical
mechanics.  What we have obtained is a problem of an atomic site
embedded into the fermionic bath: in many-body physics this system
is known as multi-band Anderson Impurity Model. While we do not
have obtained an Hamiltonian that describes the mapping system, we
can easily write down the effective action $S$, so that the
problem is fully determined.
All the mathematical details, the
explicit formulas and a more detailed description of the DMFT
equations can be found in Refs. \onlinecite{kotliar06rmp78:865} or
\onlinecite{held07ap56:829}.

The Anderson Impurity Model has been widely
studied in the many-body literature and its solution can be obtained
through many different techniques, usually named ``solvers'' in
the DMFT community. At the present time no solver has succeeded to
become the ``standard approach'' of the LDA+DMFT scheme, but the
technique to be used is every time decided with respect to the
strength of the correlations and the degree of accuracy desired.
In the case of systematic simulations, as for example the total
energy calculations reported in this paper, another important
factor to consider is the numerical efficiency. In both the
implementations of the LDA+DMFT scheme discussed here the
spin-orbit spin-polarized $T$-matrix fluctuation-exchange (so
SPTF\cite{pourovskii05prbl72:115106,katsnelson02epjb30:9}) solver
has been used, being reliable and efficient for moderate strength
of the correlations ($U \lesssim W/2$ where $W$ is the bandwidth of the
localized orbitals).

Once the effective impurity problem has
been solved and a self-energy $\hat{\Sigma}_{\vvR}(z)$ has been
obtained, there is an apparent change of the number of particles, so that the
chemical potential $\mu$ has to be updated. Furthermore a new
electronic bath $\hat{\mathcal{G}}_0^{-1}(\vvR,z)$ is defined through
the inverse Dyson equation: \be \label{eq:g0def}
\hat{\mathcal{G}}_0^{-1}(\vvR,z) = \hat{G}_{\vvR}^{-1}(z) +
\hat{\Sigma}_{\vvR}(z). \ee Now we can iterate the procedure described
above until convergence of the self-energy and the number of
particles (or chemical potential). This is the basic DMFT cycle
and is schematically reproduced in Figure 1.
\begin{figure}
\includegraphics[scale=0.2]{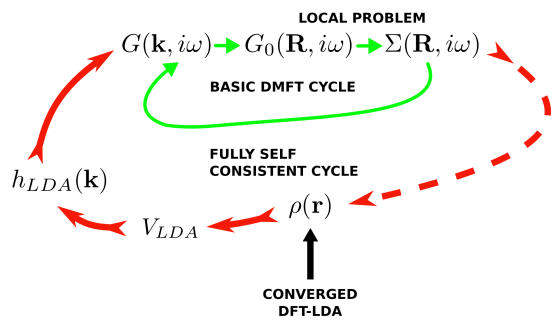}
\caption{\label{figure_1} (Color on-line) Schematic representation of the iterative procedure to follow in the LDA+DMFT scheme. As a first step the DFT-LDA problem is solved and a ground state electron density $\rho(\vvr)$ is obtained. From $\rho(\vvr)$ we can extract the matrix elements of the single-particle LDA Hamiltonian, and then build the one-electron Green's function ${G}(\vk,i \omega_n)$ at the Matsubara frequencies $i\omega_n$. Now the basic DMFT cycle starts: the Green's function ${G}(\vk,i \omega_n)$ is projected onto the correlated orbitals, defining the bath Green's function ${\mathcal{G}}_0^{-1}(\vvR,i\omega_n)$ of the Anderson Impurity Model by means of the Eq. (\ref{eq:g0def}). The solution of the local problem through one of the available ``solvers'' leads to a  self-energy function $\Sigma_{\vvR}(i\omega_n)$. After a back projection to the LDA basis set, a new one electron Green's function ${G}(\vk,i \omega_n)$ and a new chemical potential $\mu$ are calculated. The procedure is repeated iteratively until convergence in the self-energy and the chemical potential. Once the convergence of the basic DMFT cycle has been reached, a new   electron density $\rho(\vvr)$ can be calculated from ${G}(\vk,i \omega_n)$. This is the fully self-consistent cycle and should be continued until convergence in $\rho(\vvr)$.}
\end{figure}
From the same Figure, we can also notice that, if the correlations
are strong, the differences in the population of the Kohn-Sham
orbitals lead to a new electron-density $\rho(\vvr)$. In this
``full self-consistent cycle'' also the  convergence of
$\rho(\vvr)$ has to be reached. In the present paper two
implementations of the LDA+DMFT scheme are used: the first one
works only within the basic DMFT cycle, while the other one uses the
fully self-consistent cycle. There are mainly two reasons why we
have compared these two different implementations. First of all we
want to study 3$d$ states: they are not extremely localized and
consequently the effective Coulomb repulsions between them are not
very strong. Then we can reasonably suppose that the changes in
the electron-density are small, and a measure of this is given by
the comparison of the two different codes. In second place one of
the aims of our study is to investigate which numerical precision can be
obtained for the LDA+DMFT scheme, also in comparison to the modern
DFT packages, so to allow reliable calculation of sensitive
quantities as equilibrium atomic volume and bulk modulus. With
respect to this issue, we should stress that implementing full
self-consistency over the charge density is a very delicate task
that can bring additional numerical errors. Moreover the
computational effort can rise considerably, limiting the
applications of the LDA+DMFT scheme to systems with a few atoms
per unit cell.

Before presenting the total-energy functional adopted in the
LDA+DMFT scheme, a final remark has to be made. Since the LDA+U
Hamiltonian is constructed with an additional term that is already
contained in the original electronic Hamiltonian, we have to
remove from the self-energy those contributions already calculated
in the LDA. Unfortunately, there is no way to establish exactly a
correspondence between approximations within the density
functional and the Green's function (Baym-Kadanoff) functional,
then we have simply to ``guess'' which diagrammatic contributions
are included and which ones are not. For treating metals the most
common choice of the ``double counting'' correction is  the static
part of the
self-energy\cite{katsnelson02epjb30:9,braun06prl97:227601}. In the
present paper we adopt the double counting of
Ref. \onlinecite{chadov08epl82:37001}, i.e. we treat the static
contribution to the self-energy as in the LDA+U method with
around mean-field (AMF) double counting, while the other
contributions to the self energies become
\be \label{doublecounting}
\Sigma_{\xi_1,\xi_2}(z) = \Sigma_{\xi_1,\xi_2}(z) -
 \delta_{\xi_1,\xi_2} \langle{ \Sigma(0) }\rangle \: , \ee
where the average has to be determined over the orbital indices
separately per spin channel. This choice is due to the fact
that the LDA exchange-correlation potential is an
orbitally averaged quantity and has proven to be very succesfull in
describing the transition metals.

\section{Total energy functional}
In the previous section we have presented the equations that
define the LDA+DMFT scheme in terms of local problem and
self-consistent bath. These equations can be obtained with many
different techniques\cite{georges96rmp68:13}, but in perspective
of total energy calculations we have already adopted the point of
view of the spectral density-functional theory of Savrasov and
Kotliar. In a series of
papers\cite{savrasov04prb69:245101,kotliar06rmp78:865} they have
introduced a functional of both the total electron density
$\rho(\vvr)$ and the local Green's function $G_{\vvR}(z)$ for the
correlated orbitals. It is important to emphasize that these
quantities are independent, in the sense that they cannot be reconstructed
from each other. Furthermore notice that, in this framework, the
arbitrariness of the basis set of the correlated orbitals is
contained in $G_{\vvR}(z)$. Following standard methods of
quantum-field  theory the functional is constructed introducing
source terms for $\rho(\vvr)$ and $G_{\vvR}(z)$; then the variational
procedure is applied to the functional with respect to these sources.
Without presenting the  mathematical details (see references
above), we obtain the following expression for the
zero-temperature limit of the total energy: \be \label{eq:tot_en}
E=E_{LDA}\left[{\rho(\vvr) }\right] - \sideset{}' \sum_{\vk \nu
}{\varepsilon_{\vk \nu}} + \Tr[{\hat{H}_{LDA}\hat{G}}]
+ \langle{ \hat{H}_U }\rangle \ee where $\hat{H}_U$ indicates the two-particle
term in the LDA+U Hamiltonian (\ref{eq:hlpu}), and the primed
sum is over
the occupied states. Here and in the following the symbol $\Tr{}$
indicates the one-electron trace for a generic representation and
the sum over the Matsubara frequencies $i\omega$ for finite temperature
many-body formalism. We assume that the temperature effects can be
taken into account only via summation over the Matsubara
frequencies and in the DFT part only weak temperature dependence
via the Fermi distribution function is taken into
account\cite{jarborlg97rpp60:1305}. This corresponds to neglect
the temperature dependence of the exchange-correlation potential
and it is a standard procedure in electronic structure calculations
of real materials. These effects are irrelevant for the cases
under consideration
where the main temperature dependence is due to spin
fluctuations\cite{lichtenstein01prl87:067205}.

We notice that the total energy within the LDA+DMFT scheme is not
simply the expectation value of this Hamiltonian, but it consists
of several terms, in analogy to the expressions of the usual DFT.
The first term $E_{LDA}\left[{\rho(\vvr) }\right]$ contains four
different contributions, namely the ones due to the external
potential, the Hartree potential, the exchange-correlation
potential and the sum of the Kohn-Sham eigenvalues.
However in the spectral density functional theory the Kohn-Sham
eigenvalues should be re-occupied with respect to the description given by the total Green's function. Then we should remove the bare Kohn-Sham eigenvalues sum (second term of
Eq. (\ref{eq:tot_en})) and substitute it with
$\Tr{[{\hat{H}_{LDA}\hat{G}}]}$ (third term).
Moreover notice that $E_{LDA}\left[{\rho(\vvr) }\right]$ depends only on the total electron density, so it does not need to be recalculated if the basic DMFT cycle is applied. In the case of the fully self-consistent cycle, the calculation is straightforward, since it uses the same LDA-DFT machinery. This point will be analyzed in more details in the section concerning the FP-KKR implementation.

 Finally we can evaluate $\langle{
\hat{H}_U }\rangle$ through the so-called Galitskii-Migdal
formula\cite{galitskii58jetp7:96,fetter-walecka03}, an elegant way
to rewrite the expectation value of a two-particle operator in
terms of a one-particle operator as the Green's function. This
formula is based on the fact that for an Hamiltonian
\mbox{$\hat{H}=\hat{H}_0+\hat{H}_U$},
i.e. the same form of the Hamiltonian (\ref{eq:hlpu}), the
equation of motion of the Green's function states that \be
\langle{ \frac{\partial}{\partial \tau} \hat{G}(\tau)}\rangle = \langle{
\hat{H}_0 }\rangle + 2\langle{ \hat{H}_U }\rangle \ee where $\tau$ is the
imaginary time for the finite temperature formalism (the
formulation for real times and $T=0$ is completely equivalent).
Using the Fourier transform with respect to $\tau$,
we can move to the energy domain
\be \langle{
\frac{\partial}{\partial \tau} \hat{G}(\tau)}\rangle =
\Tr{[{\omega\hat{G}(\omega)}]} . \ee Furthermore from the
definition of the Green's function $ [{\omega \hat{\mathbf{1}} -
\hat{H_0} - \hat{\Sigma}(\omega) }]\hat{G}(\omega) =
\hat{\mathbf{1}} $, we can rewrite the expression above in terms
of more manageable operators \be \label{eq:gmequ}
\Tr{[{\omega\hat{G}(\omega)}]} =
\Tr{[{\hat{\Sigma}(\omega)\hat{G}(\omega)}]} +
\Tr{[{\hat{H}_0\hat{G}(\omega)}]} . \ee Then the
Galitskii-Migdal formula becomes \be \label{eq:gm} \langle{ \hat{H}_U
}\rangle = \frac12  \Tr{[{\hat{\Sigma}\hat{G}}]} \ee

\section{Implementation in FP-LMTO}
We have implemented the total-energy algorithm of the previous
section in the recently developed LDA+DMFT
code\cite{grechnev07prb76:035107}, based on the full-potential
linear muffin-tin orbital (FP-LMTO) method code presented in
Ref.\onlinecite{wills:fp-lmto} and well-known to give accurate
description of many solids within LDA. The full-potential
character of the program makes it very attractive for open
structures and surfaces, and in fact the first applications of our
code were focused on these systems\cite{grechnev07prb76:035107}.
Furthermore the use of a small number of basis functions as used
within the LMTO method
is particularly efficient for calculating the Green's functions,
since they require inversions of a matrix in the LDA basis set for
each Matsubara frequency and $\vk$ point. While we do not want to
give a complete survey of the equations involved in the FP-LMTO
code, the study of the total energy problem forced us to develop a
more sophisticated method to calculate the number of electrons for
the given chemical potential. In this implementation two basis
sets are used: the already mentioned set of the correlated
orbitals $\lfi \ket{\vvR, \xi} \rfi$ and the set of the LDA basis
functions $\lfi \ket{\vk, \chi} \rfi$. The steps of the LDA+DMFT
scheme that imply moving from the local problem to the lattice
problem require transformations back and forth between these two
basis sets. Furthermore we should mention that the set $\lfi
\ket{\vk, \chi} \rfi$ is neither normalized nor orthogonal and
then the basic algebraic operations involve an overlap matrix \be
S(\vk)_{\chi_1, \chi_2} = \braket{\vk,\chi_1|\vk,\chi_2} \ee and
its inverse $S^{-1}$, since the dual basis set of $\lfi \ket{\vk,
\chi} \rfi$ does not coincide with the set itself. The number of
electrons in the lattice problem is calculated with the LDA basis
set and becomes \be \label{eq:number_of_el} N = T \sum_{i
\omega_n} \sum_{\vk} \sum_{\chi_1,\chi_2} S(\vk)_{\chi_2,\chi_1}
{G}(\vk,i \omega_n)_{\chi_1,\chi_2} \:, \ee where \be
\label{eq:gf_to_decompose} {G}(\vk,i
\omega_n)_{\chi_1,\chi_2}=\braket{\vk,\chi_1| \hat{G}(i \omega_n)
| \vk,\chi_2} . \ee The sum over the Matsubara poles should
include infinite negative and positive frequencies, but obviously
in a computational scheme the number of frequencies can only be
finite and then a cut-off value $\omega_{max}$ needs to be chosen.
Unfortunately, as it is clear from the definition (\ref{eq:gf}),
the Green's functions decay slowly with the energy and then a
reliable determination of the number of particles would require a
huge cut-off. There are two ways to reach this cut-off: increasing
the number of Matsubara frequencies or increasing the spacing
between them, proportional to the temperature $T$. None of them is
a good solution. The former would imply too big numerical effort
(there is an inversion of a matrix with the size of
the LDA basis set for every Matsubara frequency and every $\vk$
point), while the latter would lead us too far from the $T=0$
ground-state. In Ref. \onlinecite{grechnev07prb76:035107} the
problem of the long-decaying tails of the Green's function was
solved in a rather rudimental way, given that the paper was
focused on the spectral properties, which are not as sensitive as
the ground state properties to the numerical details. In the
present paper, conversely, we follow the elegant procedure
used in the LDA+DMFT
calculations\cite{kotliar06rmp78:865,pourovskii07prb76:235101}
and adapted to
our non-orthonormal basis set. The idea is to  decompose the
calculated Green's function (\ref{eq:gf_to_decompose}) as \be
\label{eq:gf_decomposed} {G}(\vk,i \omega_n)_{\chi_1,\chi_2} =
{G}(\vk,i \omega_n)_{\chi_1,\chi_2}^{num} + {G}(\vk,i
\omega_n)_{\chi_1,\chi_2}^{an} , \ee where ${G}(\vk,i
\omega_n)_{\chi_1,\chi_2}^{an}$ is a analytical function that we
chose to fit the high-frequency behavior of ${G}(\vk,i
\omega_n)_{\chi_1,\chi_2}$: \be
\sum_{i\omega_n}^{(\omega_n>\omega_{max})} \left[{{G}(\vk,i
\omega_n)_{\chi_1,\chi_2}-{G}(\vk,i
\omega_n)_{\chi_1,\chi_2}^{an}}\right] =0 . \ee On the other hand
the numerical part is defined as the difference between the
calculated function and the analytical function \be {G}(\vk,i
\omega_n)_{\chi_1,\chi_2}^{num} \equiv {G}(\vk,i
\omega_n)_{\chi_1,\chi_2}-{G}(\vk,i \omega_n)_{\chi_1,\chi_2}^{an}
, \ee and, if ${G}(\vk,i \omega_n)_{\chi_1,\chi_2}^{an}$ has been
chosen wisely, is negligible for $\omega_n > \omega_{max}$.

The new problem is to determine $G^{an}$. Starting from the
definition (\ref{eq:gf}), we can rewrite the matrix element
(\ref{eq:gf_to_decompose}) as \be {G}(\vk,i
\omega_n)_{\chi_1,\chi_2}=\Braket{\vk,\chi_1| \bigl[{
i\omega_n-\hat{A}_{\vk}(i\omega_n) }\bigr]^{-1} \bigg| \vk,\chi_2}
\ee
where we have defined the new operator
\be
\label{eq:an_operator}
\hat{A}_{\vk}(i\omega_n) \equiv \mu \hat{\mathbf{1}} -  \hat{h}_{LDA} -  \hat{\Sigma}(i\omega_n) \:  .
\ee

Let's consider $ \hat{\Sigma}(i\omega_n) =0$ corresponding to the
first iteration of the LDA+DMFT cycle. In this case the operator
(\ref{eq:an_operator}) does not depend on the Matsubara
frequencies and is Hermitian; consequently it has real eigenvalues
$\lambda^{\vk}_m$ and the eigenvectors ${\ket{X^{\vk}_m}}$ can be
chosen to form an orthonormal set. By expanding
$\hat{A}_{\vk}(i\omega_n)$ in its spectral representation, the
Eq. (\ref{eq:gf_decomposed}) becomes
\begin{multline}
\label{eq:gf_spectralized} {G}(\vk,i \omega_n)_{\chi_1,\chi_2} =
{G}(\vk,i \omega_n)_{\chi_1,\chi_2}^{num} + \\ {+
\sum_m\frac{\braket{\vk,\chi_1|X^{\vk}_m} \braket{X^{\vk}_m |
\vk,\chi_2}}{i\omega_n-\lambda^{\vk}_m}} .
\end{multline}
We have finally reduced the
original sum to two terms that we can calculate with high
precision. The numerical term is simply calculated as a sum for
positive frequencies up to $\omega_{max}$. The sum for negative
frequencies is obtained using the symmetry of the Green's function
\be {G}(\vk, - i \omega_n)_{\chi_1,\chi_2} = [{G}(-\vk,i
\omega_n)_{\chi_2,\chi_1}]^* \ee
The analytical term can be summed through standard many-body
techniques:
\begin{multline}
\sum_{i\omega_n}{\sum_m\frac{\braket{\vk,\chi_1|X^{\vk}_m}
\braket{X^{\vk}_m | \vk,\chi_2}}{i\omega_n-\lambda^{\vk}_m}} = \\
\sum_m\frac{\braket{\vk,\chi_1|X^{\vk}_m} \braket{X^{\vk}_m |
\vk,\chi_2}}{1+e^{\beta\lambda^{\vk}_m}} .
\end{multline}
In comparison with Ref. \onlinecite{pourovskii07prb76:235101}
finding eigenvalues and eigenvectors of $\hat{A}_{\vk}$ is slightly more
cumbersome here: due to the non-orthonormality of the basis set we
have to solve a generalized eigenvalue problem. However using the
fact that the overlap matrix is positive definite, through
Cholesky decomposition\cite{numerical_recipes}
of $S$ the problem can be reduced to a usual eigenvalue problem
through a few algebraic operations.

When the DMFT self-energy assumes finite values, the operator
$\hat{A}_{\vk}(i\omega_n)$ is different at every Matsubara
frequency, and then we need to use some approximation. Luckily in
many-body theory  the analytical properties of the self-energy
operator are the same as for the Green's function. Therefore
we can assume the following asymptotic behavior for high
frequencies\cite{pourovskii07prb76:235101}: \be
\label{eq:asymp_for_sigma} \hat{\Sigma}(i\omega_n) \sim
\hat{\Sigma}^{stat} + \frac{\hat{\Sigma}^{asym}}{i\omega} , \ee
where $\hat{\Sigma}^{stat}$ and $\hat{\Sigma}^{asym}$ are obtained
from the real and imaginary part of $\hat{\Sigma}$ at the cut-off
value $\omega_{max}$. While a higher $\omega_{max}$ will always
give a better fit, the real part of the self-energy converges to
$\hat{\Sigma}^{stat}$ as $1/\omega^2$, and then we do not need a
very high cut-off. Furthermore for our purposes of evaluating the
frequency sum in Eq. (\ref{eq:number_of_el}), we can keep
only the dominant term $\hat{\Sigma}^{stat}$, and
$\hat{\Sigma}^{asym}$  turns to be unimportant. Again the operator
(\ref{eq:an_operator}) becomes Hermitian and independent on the
Matsubara frequencies, so that the same procedure described above
can be applied.

The implementation of this algorithm in the FP-LMTO code improved
the precision in the determination of the number of particles by
about two orders of magnitude in the worst cases (corresponding to
a large number of LDA basis functions that increases the numerical
error on the eigenvectors). The method used in
Ref. \onlinecite{grechnev07prb76:035107} was rather
similar to the one presented above, but had a much simpler 
implementation. Instead of considering the asymptotic behavior of
every Green's function in Eq. (\ref{eq:number_of_el}), the sum over
the intermediate indices $\chi_1$,$\chi_2$ and $\vk$ was done, and
then the asymptotic behavior of the resulting function was
considered. While this approximation can appear too crude, the
precision on the number of particles is about $10^{-3}$ particles
for every electron involved in the problem. On the other hand it
was computationally very efficient, since the generalized
eigenvalue problem was reduced to the determination of a pure
number.

After having improved the precision in the determination of the
number of particles, we can pass to the implementation of the
total energy formula (\ref{eq:tot_en}). As we have already seen
the first two terms can be obtained from the density-functional
part of the LDA+DMFT scheme. The third term, corresponding to the
reoccupation of the Kohn-Sham orbitals, requires again the
evaluation of a sum over all the Matsubara frequencies \be
\label{eq:single_energy} \Tr{[{\hat{H}_{LDA}\hat{G}}]} =
T \sum_{i \omega_n} \sum_{\vk} \sum_{\chi_1,\chi_2}
H_{LDA}(\vk)_{\chi_2,\chi_1}  {G}(\vk,i \omega_n)_{\chi_1,\chi_2}
. \ee Besides the presence of different matrix elements, 
Eq. (\ref{eq:single_energy}) is completely analogous to 
Eq. (\ref{eq:number_of_el}), therefore the sum can be done by
applying the same procedure used above. Finally we have to evaluate
the Galitskii-Migdal contribution $\langle{ \hat{H}_U }\rangle$. Given
that in the LDA+DMFT scheme the self-energy is local, the trace in
Eq. (\ref{eq:gm}) can be restricted to the correlated
orbitals. Furthermore, using the fact that in the SPTF solver
we work with quantities in both the frequency and (imaginary) time
domains, we can express the trace in terms of the complex Fourier
transforms. For this purpose, it is most convenient to separate the
static and the dynamic parts of the self-energy. Analogously to
Eq. (\ref{eq:asymp_for_sigma}), we have \be \hat{\Sigma}(i\omega_n) =
\hat{\Sigma}^{stat} + \hat{\Sigma}(i\omega_n)^{dyn} \:. \ee 
However now no fitting is necessary:
once $\hat{\Sigma}^{stat}$ is determined,
$\hat{\Sigma}(i\omega_n)^{dyn}$ contains all the differences with
the calculated function $\hat{\Sigma}(i\omega_n)$. We can then
write \be \label{eq:gm_trace} \langle{ \hat{H}_U }\rangle = \frac12 T
\sum_{i\omega_n} {\sum_{\xi_1,\xi_2}
{[{\Sigma_{\xi_1,\xi_2}^{stat} +
\Sigma(i\omega_n)_{\xi_1,\xi_2}^{dyn}}]
G(i\omega_n)_{\xi_2,\xi_1}}} . \ee The first term at right hand
side can be easily Fourier transformed and reduced in terms of
occupations of the local orbitals \be
n_{\xi_1,\xi_2}=G(\tau=0^{-})_{\xi_1,\xi_2} ; \ee the second term
requires the evaluation of the Fourier transform of a product,
leading to a convolution. In summary we can express
Eq. (\ref{eq:gm_trace}) as
\begin{multline}
\label{eq:gm_in_flex}
\langle{ \hat{H}_U }\rangle = \frac12 \sum_{\xi_1,\xi_2}{ [{\Sigma_{\xi_1,\xi_2}^{stat}n_{\xi_2,\xi_1}}} + \\
+ \int_0^{\beta}{d\tau \: {\Sigma(\tau)_{\xi_1,\xi_2}^{dyn} G(-\tau)_{\xi_2,\xi_1}}}] .
\end{multline}

\section{Implementation in FP-KKR}\label{ImplKKR}
The same total energy algorithm of the previous sections was implemented
in the FP-KKR code described in Ref. \onlinecite{chadov08epl82:37001},
being an extension to the full-potential case of the earlier ASA
implementation\cite{minar05prb72:045125}. Besides the advantage of
being one of the very few fully-self consistent implementations of
the LDA+DMFT scheme, the formalism on which the FP-KKR code relies
makes it particularly attractive to study complex problems as 
orbital polarizations\cite{chadov08epl82:37001}, photoemission
spectroscopy through the one-step model\cite{braun06prl97:227601}, or
disordered alloys systems through CPA\cite{sipr08prb}.
As drawback to the flexibility of LDA+DMFT, in FP-KKR we have a high
computational cost that can make it inconvenient to
perform extensive simulations,
e.g. determination of total energy curves as functions of 
the crystal parameters, compared to other simpler methods.

Without presenting a complete survey of the equations involved in the
FP-KKR method, we should mention that it is based on the multiple
scattering theory which allows to decompose the total one-particle
Green's function  into the single scattering matrix $t^{\vvR}(\epsilon)$ 
which contains
the information about each single scatterer $\vvR$
(i.e. atomic site) and the structure constants matrix ${\cal G}^{\vvR\vvR'}(\epsilon)$
which contains the information about the geometrical arrangement of the
scatterers in a solid. All the ingredients are calculated in the 
basis of the four-component
energy-dependent regular and irregular solutions of the 
relativistic Kohn-Sham-Dirac equations\cite{macdonald79jpc12:2977,
ramana79jpc12:L845}. Corresponding DMFT self-energy is
included into corresponding quasiparticle Dirac equation, e.g. in
contrast to the other LMTO based LDA+DMFT implementations the
correlation effects are included directly into the single site $t$ matrix
as well as into the wave functions at the same time.

In order to construct the bath Green's function needed as
an input for the DMFT solver, the localized Green's
function is calculated by projecting the total Green's function
onto the correlated atomic site. The multiple scattering
formalism provides the natural choice of the projectors
which are nothing else as the regular single-site 
solutions of the Kohn-Sham-Dirac equations. The projection functions
are taken at the fixed energy, which corresponds to the center of
mass of the band and is recalculated at each iteration.

\begin{figure}
\includegraphics[scale=0.3]{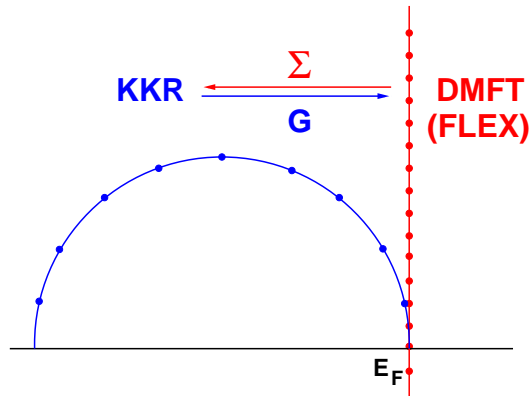}
\caption{\label{figure_2} (Color on-line) Illustration of the KKR+DMFT
  scheme: blue semicircle is the complex energy path used by KKR to
  calculate the Green's function. After the bath Green's function $G$
  is obtained, it is analytically continued onto the imaginary axis
  (red) to calculate the self-energy  via the SPTF impurity solver. The
  latter is analytically extrapolated back to the semicircle.}
\end{figure}

Different from FP-LMTO, the FP-KKR code works with the Green's functions
from the beginning, so that the merging between the LDA part and the DMFT
part in the LDA+DMFT scheme does not require a change of representation
of the electronic states. In practice,
as shown in Figure \ref{figure_2} the LDA Green's function
is evaluated on a semicircular contour in the complex plane, but the
SPTF solver works with the Green's functions on the Matsubara frequencies.
The analytical continuation of the self-energy from the Matsubara axis to
the semicircular contour is done through the Pad\'{e} approximants method,
and this could introduce small numerical errors. While no problem was
observed in all the previous studies for which this code has been used,
we have considered that the determination of the energetic landscape
requires more precision than spectral properties. For that reason
we have checked this point carefully. As expected
we have found a small numerical noise, but in practice its effects
on the ground state properties of the transion metals studied here
were negligible.

In FP-KKR the total energy functional (\ref{eq:tot_en}) can be rewritten in
a slightly different form. In the LDA contribution to the LDA+DMFT total
energy we can explicitate the standard terms that are going to be summed in
density functional theory. We have
\begin{multline}
\label{eq:KKR_LDA_en}
E_{LDA}\left[{\rho(\vvr) }\right]=\int^{\epsilon_F}{d\epsilon \: \epsilon N_R(\epsilon)} - \int_{S_R}{d\vvr \: \rho(\vvr) V_R(\vvr)} +\\
+ \int_{S_R}{d\vvr \: \rho(\vvr) \left\{{\int_{S_R}{d\vvr' \: \frac{\rho(\vvr')}{|\vvr - \vvr'|}}-\frac{2Z_R}{|\vvr|} -\varepsilon_{xc}[\rho(\vvr)]  }\right\}}
\end{multline}
where $R$ is a given region of the space defined in the unit cell and
$N_R$ is the number of electrons in the space $R$. This occupation
number is obtained from the unperturbed LDA Green's functions, that is
from the information specified by $\rho(\vvr)$.
This is the reason why the functional dependence of the energy
above is restricted to the only electron density.
If we calculate the number of electrons $N_R$ with the full DMFT
Green's functions, we obtain that the first term of the functional
(\ref{eq:KKR_LDA_en}), the so-called ``band energy'', becomes
exactly the term at left hand side of Eq. (\ref{eq:gmequ}).
Then, renaming the quantity (\ref{eq:KKR_LDA_en}) calculated with
this new occupation as $E_{LDA}\left[{\rho(\vvr), G(\omega)}\right]$,
it is
straightforward to rewrite the functional (\ref{eq:tot_en}) as
\be
\label{eq:kkr_functional}
E_{LDA+DMFT}=E_{LDA}\left[{\rho(\vvr), G(\omega)}\right]
 -  \langle{ \hat{H}_U
}\rangle 
\ee
where the Galitskii-Migdal term has to be subtracted since
it is already accounted for twice within the band energy.

The evaluation of the formula (\ref{eq:kkr_functional}) requires only
the calculation of the Galitskii-Migdal energy (\ref{eq:gm}), since
the band energy results from the DFT part of the FP-KKR code. While
it could be simpler to evaluate the Galitskii-Migdal correction
directly on the local problem through the formula (\ref{eq:gm_in_flex}),
we prefer to work again on the semicircular complex contour, 
retaining to the same formalism for both the contributions to the total
energy. Then we calculate
\begin{eqnarray}
  \label{eq:KKR-gm}
  \langle{ \hat{H}_U }\rangle =-\frac{1}{2\pi}{\rm Im} \sum_{\xi_1\xi_2}\int{dz}\,\Sigma_{\xi_1\xi_2}(z)G_{\xi_2\xi_1}(z)\,.
\end{eqnarray}
The integration is performed over the contour starting close to the real
energy axis at the bottom of a valence band and ending at the Fermi
energy. It turned out that this procedure is numerically more stable
than evaluation of Galitskii-Migdal correction using integration over 
the Matsubara frequencies.

\section{fcc Ni}

Bulk fcc Ni is a sort of standard test-case for every approach
to strongly correlated materials. For this reason it has been
chosen
as first application for the implementations presented above. The
interest of the DMFT community in Ni
started\cite{lichtenstein01prl87:067205} with the explanation of
the famous 6 eV satellite observed in photoemission experiments,
but missing in all DFT calculations. Afterwards spectral
properties of bulk Ni were studied through different LDA+DMFT
implementations\cite{katsnelson02epjb30:9, grechnev07prb76:035107}
and also through the GW+DMFT
calculations\cite{biermann03prl90:086402}. All these studies
confirmed the correlated nature of the Ni satellite. Furthermore
recent LDA+DMFT based calculation of the one-step model
photoemission spectrum showed a very good quantitative agreement
with experimental photoemission data\cite{braun06prl97:227601}.
Along with these spectral features, the LDA+DMFT method has been
applied to the finite-temperature
magnetism\cite{lichtenstein01prl87:067205} of Ni, showing the
existence of local moments (unordered above the Curie
temperature), i.e. another clear sign of strong correlation.
Nevertheless we should consider that the DFT scheme is not focused
on the excitation spectrum, but mainly on the electron density.
Given the the LDA+DMFT scheme and the Hamiltonian (\ref{eq:hlpu})
are explicitly build for the correct description of the low-energy
excitations, it appears natural that this scheme performs
convincingly better than simple density functional theory.
Conversely DFT gives a reasonable description of all ground
state properties of Ni and the agreement with the experimental
data becomes almost perfect if GGA is
used\cite{guo00cjp38:949,cerny03prb67:035116,cerny07mse462:432}.
Moreover, in contrast with the other late transition metals, the
inclusion of the spin polarization in the calculations for
fcc Ni is not strictly necessary, surely due to the small
magnetic moment ($\mu \simeq 0.6$)
acquired\cite{cerny03prb67:035116} at the equilibrium structure.
Finally, a recent accurate study of the orbital and spin
polarization of the late transition
metals\cite{chadov08epl82:37001} emphasized that the DMFT
corrections to the DFT-LDA values for Ni are really minor, while
still improving the description of the material.

With reference to the previous arguments, it appears necessary
to clarify the reasons behind our interest in the
energetics of fcc Ni, where the correlation effects are expected
to have a moderate role.
\begin{figure}
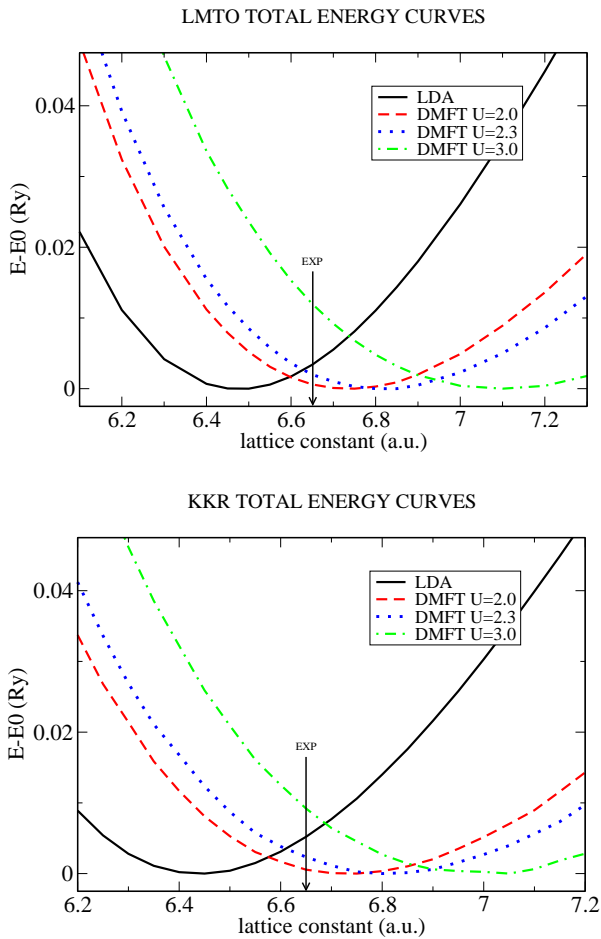

\includegraphics[scale=0.3]{figure_3_top.eps}
\vspace{0.5cm}
\hfill

\includegraphics[scale=0.3]{figure_3_bot.eps}
\caption{\label{figure_3} (Color on-line) Energy vs lattice constant curves for fcc Ni in the DFT-LDA scheme and in the LDA+DMFT scheme based on the FP-LMTO (top) and FP-KKR method (bottom). The zero of the energy of each curve is set to its own minimum value $E_0$ and three chosen values of $U$ are presented ($T=400 \text{ K}$). The experimental value of the lattice constant is indicated by the arrow.}
\end{figure}
First of all it is important to complete the picture
outlined above: excitation spectrum,
magnetism, photoemission spectrum, surfaces, orbital polarization
and now ground state properties. This study can help in
understanding how correlated fcc Ni is\cite{held07ap56:829} and
which deficiencies of the DFT-LDA technique are due to a
single-particle approximation of the exchange-correlation
potential and which ones are due to the intrinsic meaning of the
Kohn-Sham quasiparticles as fictitious excitations.  In second
place Ni represents a good test-case to prove the ability of
the LDA+DMFT scheme to catch moderate correlation effects
in a real material. In fact we know that the LDA+DMFT scheme
relies mainly on two different approximations: finite number
of nearest neighbors (due to the locality of the self-energy)
and non-exact solver. Therefore it is interesting to check
how dominant are the errors connected to these approximations
for effects that are expected to be rather small.
Furthermore a third important question concerns the role of the
full self-consistency in the DMFT cycle. Previous
studies\cite{pourovskii07prb76:235101} for $\text{Ce}_2\text{O}_3$
and $\gamma \mbox{-} \text{Ce}$ have shown, quite surprisingly,
small differences between the ground state properties for the
basic and fully self-consistent DMFT
cycles. Given that these systems involve valence electrons much
more localized than the ones of Ni, in our case we expect
negligible differences, at least in the range of ``acceptable''
Hubbard $U$. This would represent a further validation of our
previous studies\cite{grechnev07prb76:035107} of bulk and surface
Ni, founded on the basic LDA+DMFT cycle, limiting the necessity of the
full cycle to the most sensitive quantities like photoemission
spectrum\cite{braun06prl97:227601} and magneto-optical
properties\cite{perlov03prb68:245112}. Finally, a last question
investigated for fcc Ni concerns the compatibility between
different implementations: can different codes with different
choices of the correlated orbitals give comparable results?

To investigate all the various points outlined in the previous paragraph,
we performed LDA+DMFT
simulations of fcc Ni for various lattice constants starting from
$a=6.2 \text{ a.u.}$ and up to $a=7.4 \text{ a.u.}$. 
We treated 3d, 4s and 4p electrons as valence electrons.
For the FP-LMTO simulations, the description of the valence
electrons in the interstitial space between the muffin tin spheres
requires LMT-Orbitals with different tail energies, whose number
depends on the degree of localization-delocalization of the
electrons: three tails were used for $4s$ and $4p$ electrons, only
two tails for $3d$ electrons. The set of the correlated orbitals
was build from the LMT-Orbitals, considering only the part
contained into the muffin-tin sphere at a given linearization
energy, the so-called ``head of the
LMTO''\cite{grechnev07prb76:035107}. Convergence on the total
energy with respect to the $\vk$-mesh lead to a minimum number of
4913 $\vk$-points used in the three dimensional Brillouin zone. A
simulation has been considered converged if the energy difference
for two consecutive iterations has been at least smaller than $0.1
\text{ meV}$. 
As far as possible same settings were used for the FP-KKR simulations with the
exception of the set of correlated orbitals (see
Sec. \ref{ImplKKR}). KKR total energies are very sensitive to the
angular momentum expansion used for calculation. To get accurate
results we performed LDA numerical tests up to $l_{\rm max}=6$. We found that
in the case of Ni and Mn to obtain converged results we need to use at
least angular momentum expansion up to $l_{\rm max}=3$. This
cut-off was used for the more computationally demanding
LDA+DMFT calculations.

The local problem was studied for different values of $U$ in the
range between $2$ and $3 \text{ eV}$, considered acceptable from
the results of constrained LDA calculations\cite{oles84prb29:314,bandyopadhyay89prb39:3517}
and previous LDA+DMFT simulations. The temperature was
set as $T=400 \text{ K}$ and $2048$ Matsubara frequencies were
used. As for the DFT part, convergence in the LDA+DMFT total
energy was considered acceptable when the changes for subsequent
iterations were smaller than $0.1 \text{ meV}$.
\begin{table*}[t]
  \centering
  \caption{Computed values of the equilibrium atomic volume $V_0$   and the bulk modulus $B$ for the the standard LDA-DFT method and for the LDA+DMFT scheme. Different strength of the local Coulomb repulsion $U$ have been studied, at $T=400 \text{K}$. The values taken from Ref. \onlinecite{cerny03prb67:035116} are obtained by means of an ASA-LMTO code.}

  \begin{tabular}{|c|c|c|c|c|c|c|}
  \hline
   &  \hspace{0.35cm} LDA \hspace{0.35cm}  & \hspace{0.02cm} $U=2.0$ eV \hspace{0.02cm} & \hspace{0.02cm} $U=2.3$ eV \hspace{0.02cm} & \hspace{0.02cm} $U=3.0$ eV \hspace{0.02cm} & \hspace{0.35cm} GGA \hspace{0.35cm} &  \hspace{0.25cm} EXP \hspace{0.25cm}  \\
  \hline
  \hline
  \hline
  \begin{tabular}{c|l}
   & FP-LMTO \\ [1ex]
   $ V_0 (\text{a.u.}^3)$ & KKR \\ [1ex]
   & Ref. \onlinecite{cerny03prb67:035116} \\ [1ex]
  \end{tabular}
  &
     \begin{tabular}{c}
      67.88 \\ [1ex]
      66.86 \\ [1ex]
      67.71 \\ [1ex]
     \end{tabular}
   &
         \begin{tabular}{c}
          76.20 \\ [1ex]
          76.28 \\ [1ex]
           \\ [1ex]
         \end{tabular}
    &
             \begin{tabular}{c}
              79.19 \\ [1ex]
              79.02 \\ [1ex]
                \\ [1ex]
             \end{tabular}
      &
                 \begin{tabular}{c}
                   89.48 \\ [1ex]
                   85.53 \\ [1ex]
                     \\ [1ex]
                 \end{tabular}
        &
                 \begin{tabular}{c}
                   \\ [1ex]
                   \\ [1ex]
                   76.54 \\[1ex]
                 \end{tabular}
         &
           73.52
           \\
  \hline
  \hline
  \hline
  \begin{tabular}{c|l}
   & FP-LMTO \\ [1ex]
   $B(\text{GPa})\:\:$ & KKR \\ [1ex]
   & Ref. \onlinecite{cerny03prb67:035116} \\ [1ex]
  \end{tabular}
  &
     \begin{tabular}{c}
      260 \\ [1ex]
      280 \\ [1ex]
      270 \\ [1ex]
     \end{tabular}
   &
         \begin{tabular}{c}
          163 \\ [1ex]
          171 \\ [1ex]
           \\ [1ex]
         \end{tabular}
    &
             \begin{tabular}{c}
              142 \\ [1ex]
              150 \\ [1ex]
               \\ [1ex]
             \end{tabular}
      &
                 \begin{tabular}{c}
                   84 \\ [1ex]
                   132 \\ [1ex]
                    \\ [1ex]
                 \end{tabular}
      &
                 \begin{tabular}{c}
                   \\ [1ex]
                   \\ [1ex]
                   186 \\ [1ex]
                 \end{tabular}
       &
          186
            \\
   \hline
\end{tabular}
\end{table*}

At the top of Figure \ref{figure_3}, we can see the total energy curves as
functions of the lattice constant for the FP-LMTO implementation.
The curves have been shifted with respect to their minima, so it
is easier to compare them. As observed in previous
calculations\cite{cerny03prb67:035116}, in DFT-LDA the equilibrium
value of the lattice constant is  slightly ($3\%$)  underestimated
with respect to the experimental one. Looking at the curves for
the LDA+DMFT simulations, we immediately notice that the results
are strongly dependent on the value of the Hubbard $U$.
Furthermore the best result seems to be obtained for $U=2 \text{
eV}$, i.e. for a value smaller than the widely accepted $U=3
\text{ eV}$. On the other hand the curve for  $U=3 \text{ eV}$
seems to comprehend too strong correlation effects. The
explanation of these results is in the perturbative nature of the
SPTF solver, which tends to overestimate correlation effects
in fcc Ni. This was noticed since the first
implementation\cite{katsnelson02epjb30:9}, when comparison between
LDA+DMFT results with the SPTF solver and numerically exact
Quantum Monte-Carlo solver showed the best agreement for $U=2
\text{ eV}$. Furthermore in the already mentioned calculation of
the orbital polarization of Ni, it is shown that SPTF with $U=3
\text{ eV}$ gives too strong correction of the orbital
moment\cite{chadov08epl82:37001}.

On the other hand we could be tempted to think that this behavior
is increased by to the lack  of the full self-consistency in the
LDA+DMFT cycle. This doubt is removed by looking at the results for
KKR, reported at the bottom of Figure \ref{figure_3}. In fact we can barely
notice any difference with respect to the energy curves of the FP-LMTO.
It is important to empahsize how similar the presented results are,
since the arbitrariness of the LDA+U Hamiltonian
(\ref{eq:hlpu}), due to the arbitrary choice of the correlated
orbitals, is often considered as a limit of the orbital-dependent
methods.

Table I, where the equilibrium atomic volume $V_0$ and
the bulk modulus $B$ are given, allows a more quantitative
comparison of the two implementations and with previous
DFT-LDA studies of fcc Ni\cite{cerny03prb67:035116}.
These values of $V_0$ and $B$ have been computed
with polynomial fitting of the energy versus atomic volume curve
around the minimum. In addition also fitting through
Birch-Murnaghan equation of state\cite{murnaghan44,birch52} was done, leading to almost
identical results and confirming the stability of our data.

As for the total energy curves, the best results are obtained for
$U=2 \text{ eV}$, and we can see that the inclusion of local
correlation effects into the LDA results corrects both the
equilibrium atomic volume and the bulk modulus in the right way.
While this fact has enough interest by its own, we should notice
that to have more precise results on the quantitative point of
view, a more strict relation between solver, correlated orbitals
and values of $U$ is needed.  Naturally it would be interesting to
repeat those calculations with the numerically exact quantum
Monte-Carlo solver to check if better agreement with the
experiment can be obtained. Another interesting property can be
deducted from the Table I: while the equilibrium atomic volumes are
independent on the full self-consistency, the bulk modulus looks
to be more strongly influenced. As expected this discrepancy is
proportional to the strength of $U$. The simulation for
the strongest value tried, i.e. $U=3 \text{ eV}$, shows the
tendency of the FP-LMTO to underestimate the value of the bulk
modulus of fcc Ni.

\section{$\gamma$-Mn}
Mn is definitely one of the most interesting and complex
materials among pure transition metals. According to Hund's
rule, free atom possesses a large magnetic moment of $5 \mu_B$, and
the stabilization of such large magnetic moments, e.g. in Heusler
alloys, would represent a great technological advance,
suitable for many applications.

Experimentally Mn exists in four different phases. The
low-temperature low-pressure phase is the
$\alpha$-phase\cite{lawson94jap76:7049}. It has a complex cubic
structure with 58 atoms per unit cell and non-collinear
antiferromagnetic order. The local moment depends strongly on the
atomic site, varying between $3 \mu_B$ and 0, and disappears
above the Ne\'el temperature $T_N = 95 \text{ K}$. At $T=1073
\text{ K}$ there is a transition to the
$\beta$-phase\cite{okeefe77acta33:914}, a cubic structure with 20
atoms per unit cell and small magnetic moment. Between $T=1368
\text{ K}$ and $T=1406 \text{ K}$ a high-temperature
$\gamma$-phase with fcc structure appears. Interestingly this
phase can be stabilized until room temperature through the
addition of a small amount of impurities\cite{endoh71jpsj30:1614}
or as layer-by-layer deposition on
$\text{Cu}_3\text{Au}(100)$\cite{schirmer99prb60:5895,biermann04zhetfl80:714}.
Below the Ne\'el temperature, about $T_N = 540 \text{ K}$ the
$\gamma$-phase acquires an anti-ferromagnetic ground-state, which
is accompanied by tetragonal distortion into the fct
structure\cite{schirmer99prb60:5895,oguchi84jmmm46:L1}. From
$T=1406 \text{ K}$ up to the melting temperature $T_M=1517 \text{
K}$ there is a $\delta$-phase, whose structure is bcc and order is
antiferromagnetic. Finally high-pressure studies have revealed a
transition to an hcp $\epsilon$-phase\cite{fujihisa95prb52:13257}
at $165 \text{ GPa}$.

Such a rich phase diagram corresponds to an equivalently rich
history of theoretical studies (for an extended and detailed
review we redirect the reader to the Ref.
\onlinecite{hafner05prb72:144420}). Obviously these studies have
been mainly focused on the two ``simplest'' phases, $\gamma$ and
$\delta$, while the increase of the computational power achieved
in the
last ten years made the first {\itshape{ab-initio}} calculations
of $\alpha$ and $\beta$ phases
appear\cite{hobbs01jpcm13:L681,hobbs03prb68:014407,hafner03prb68:014408}.

Our main interest concerns the ground-state properties of $\gamma$-Mn
and the role of correlation effects. The description of
the electronic properties given by density-functional theory is
undoubtedly wrong for non spin-polarized LDA,
but it becomes more reasonable
if spin-polarization is
introduced\cite{moruzzi89prb39:6957,moruzzi93prb48:7665}. As for Fe,
however, LSDA does not predict the correct crystal structure, but
the ground-state of Mn results to be
hcp\cite{asada93prb47:15992}. Furthermore these strong
magneto-volume effects are reflected into an anomalously low value
of the bulk modulus\cite{moruzzi93prb48:7665}. This can be
considered as a first hint to strong correlation effects. Like for
the other transition metals, the agreement of the calculated
ground-state properties with the experimental data improves
drastically if spin-polarized GGA is used as exchange-correlation
potential\cite{eder00prb61:11492,hafner05prb72:144420}, but the
discrepancies  are still the strongest of the $3d$ series.
Furthermore, as already pointed out by
Zein\cite{zein95prb52:11813}, the anomalous properties of
Mn do not seem to depend so strongly on the magnetic phase.
In fact extrapolation of experimental data for Mn-Cu alloys to
zero content of Cu shows\cite{tsunoda84jpsj53:359} equilibrium
atomic volume and bulk modulus comparable (in a range of 10\%) to
pure $\gamma$-Mn, while doping by Cu suppresses
antiferromagnetism in $\gamma$-Mn. The situation becomes still
worse if spectral properties are considered. The only LDA+DMFT
study available on $\gamma$-Mn has
shown\cite{biermann04zhetfl80:714} that inclusion of local Coulomb
interactions is necessary for a proper description of the
excitations. Following this work, $\gamma$-Mn seems to
behave more as a strongly correlated metal at the metallic side of
Mott metal-insulator transition, than as a moderately correlated
metal with some deficiencies in the spectrum, as Ni: Hubbard
bands are formed for high energies and a quasiparticle resonance
appears around the Fermi level. To clarify the role of
correlations and the connection between correlations and magnetism
in $\gamma$-Mn we have carried out systematic LDA+DMFT
simulations. We have adopted a simple fcc crystal structure in a
layered antiferromagnetic phase AFM1, since previous
simulations showed clearly this to be the equilibrium
structure\cite{kruger96prb54:6393,eder00prb61:11492,hafner05prb72:144420}.
As already deduced in the early eighties\cite{oguchi84jmmm46:L1}, the frustration of
the AFM1 fcc structure should imply a slight (6\%) distortion of
the lattice, but this effect has not been considered here, since
its role is not so important in comparison to local Coulomb
interactions. The relation between correlation effects, frustration
and lattice distortion will be the subject of future
investigations. The lattice constants have been ranged from $a=6.0
\text{ a.u.}$ and up to $a=7.5 \text{ a.u.}$. All the other
computational details have been set as the ones used for
Ni.

The choice of the Hubbard $U$ for Mn is not trivial at all,
since this element was not studied as much as Ni. In the
previous LDA+DMFT study\cite{biermann04zhetfl80:714} it was varied
between $3\text{ eV}$ and $5\text{ eV}$ through semi-empirical
considerations. However, recent progress has been made on the
implementation of procedures to determine the parameters
describing the local Coulomb interactions {\it{ab initio}}. New
results for the $3d$ transition metals have been obtained using
the ``canonical'' constrained local density
approximation\cite{nakamura06prb74:235113} and the ``new''
constrained random-phase
approximation\cite{aryasetiawan04prb70:195104,aryasetiawan06prb74:125106}
and they locate $U$ in the range $2-4\text{ eV}$ for the whole
series, reaching maximum values for the half-filled systems. Given
that one of these simulation used a basis set very similar to ours
(head of the LMTO)\cite{aryasetiawan06prb74:125106}, for
$\gamma$-Mn we adopted  $U= 2.6\text{ eV}$ and $U=
3.0\text{ eV}$. The corresponding Stoner parameter was chosen as,
respectively, $I= 0.8\text{ eV}$ and $I= 0.9\text{ eV}$.

\begin{figure}
\includegraphics[scale=0.3]{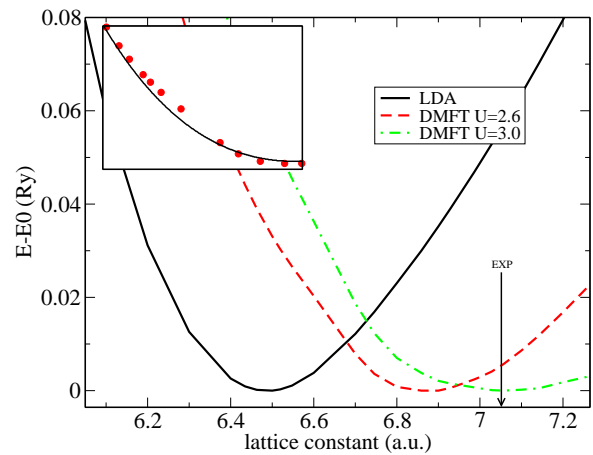}
\caption{\label{figure_4} (Color on-line) Energy versus lattice constant curves for $\gamma$-Mn in the DFT-LDA scheme and in the LDA+DMFT scheme based on the FP-LMTO method. The zero of the energy of each curve is set to its own minimum value $E_0$ and two chosen values of $U$ are presented ($T=400 \text{ K}$). The lattice constant that corresponds to the experimental atomic volume is indicated by the arrow. In the inset we can observe the total energy for LDA+DMFT simulation at $U=2.6\text{ eV}$ (big points) as function of the atomic volume compared to the standard Birch-Murnaghan equation of state (solid line). }
\end{figure}
In Figure \ref{figure_4} the total energy curves as
functions of the lattice constant for the FP-LMTO implementation
are given.
As for Ni, the curves have been shifted with respect to their
minima to obtain a better visualization.
\begin{figure}
\includegraphics[scale=0.3]{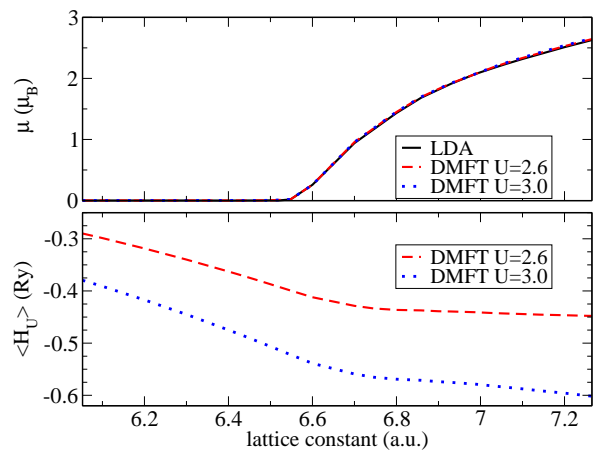}
\caption{\label{figure_5} (Color on-line) Local magnetic moment $\mu$ and Galitskii-Migdal contribution to the total energy $<\hat{H}_U>$ as function of the lattice constant for $\gamma$-Mn. While it is not observable from the picture the magnetic moment of the LDA+DMFT simulation is increased with respect to its bare LDA value. For $U=2.6\text{ eV}$ the increase in the magnetic moment is about $0.02 \: \mu_B$, while for $U=3.0 \text{ eV}$ it is about $0.03 \: \mu_B$. Interestingly no magnetic moment is created if the starting Kohn-Sham densities is   non-magnetic.}
\end{figure}
From Figure  \ref{figure_4}, we immediately notice two
interesting features in the LDA+DMFT total energy curves. 
First of all we can notice that,
by increasing the value of $U$ from zero to the
accepted effective value, the minima of the total energy curves
of the LDA+DMFT simulations gradually tend to the experimental
lattice constant. Furthermore the dependency of the results
from the strength of $U$, which have been already observed for Ni,
looks still bigger and we consider it as good indication for strong
correlations. This impression is emphasized
by another interesting feature noticeable from
Figure  \ref{figure_4}: the total energy curves do not appear
to have a perfect parabolic shape as for
usual LDA or GGA simulations, or also for the LDA+DMFT simulations
of Ni depicted in Figure \ref{figure_3}.
Instead they show a small kink for lattice constants around
$6.6\text{ a.u.}$. To make it more visible, in the inset of Figure
\ref{figure_4} the calculated data for $U=2.6\text{ eV}$ are
compared with a standard fitting through Birch-Murnaghan equation
of state. This kink is a clear sign of the strongly correlated
character of $\gamma$ Mn and reminds the one found in
LDA+DMFT total energy curves of
$\delta$-plutonium\cite{savrasov01nature410:793}. In the latter
case, there is more than just kink, there is a second minimum of
the total energy which was associated with the volume of
monoclinic $\alpha$ phase. For Mn, there is no phase transitions
with large volume jumps, like for Pu, but, instead, anomalies of
the bulk modulus in Mn-based alloys are
observed\cite{tsunoda84jpsj53:359}. It is important therefore to
analyze the origin of this kink. In Figure \ref{figure_5} magnetic
moments and Galitskii-Migdal contributions to the total energy
functional are shown. We can see that the value of the lattice
constant corresponding to our kink is a bit higher than the
critical value for which the non-zero magnetic moment appears. At
the onset of the magnetism, the competition with the local Coulomb
interactions brings a saturation of the Galitskii-Migdal energy,
which otherwise would be expected to decrease with the atomic
volume (as for example we observe for Ni).
Instead of decreasing the correlation energy, the system
responds with an increase of the magnetic moment with respect to
the bare LDA value. This change is so small that it can be barely
noticed in the upper plot of Figure \ref{figure_5}.  For
$U=2.6\text{ eV}$ the increase of the magnetic moment is about
$0.02 \: \mu_B$, while for $U=3.0 \text{ eV}$ it is about $0.03 \:
\mu_B$. 

Given that the FP-LMTO implementation is numerically less expensive than
FP-KKR, we have made extensive calculations for $\gamma$-Mn
only using the former method. A few simulations have been made also
with the FP-KKR method and the same qualitative features
reported in Figures  \ref{figure_4} and \ref{figure_5} have been
observed, stating again that for the
description of the ground state properties of $3d$ transition metals
the inclusion of the local correlation effects on the electron
density is not strictly necessary. 

A more clear picture of the physical properties of $\gamma$-Mn can be
obtained from the Table II, where equilibrium atomic volume
$V_0$, bulk modulus $B$ and magnetic moment $\mu$ for our simulations
have been compared to the experimental values and to the results
reported in Ref. \onlinecite{eder00prb61:11492}.

\begin{table*}[t]
  \centering
  \caption{Computed values of the equilibrium atomic volume $V_0$, the bulk modulus $B$ and magnetic moment $\mu$ of $\gamma$-Mn for the
standard LDA-DFT method and for the LDA+DMFT scheme. Different strengths of the local Coulomb repulsion $U$ have been studied, at $T=400 \text{K}$. The values taken from Ref. \onlinecite{eder00prb61:11492} are obtained by means of a USPP-PAW (ultrasoft pseudopotential  projector augmented plane-wave) code, and using the Murnaghan equation of state\cite{murnaghan44,birch52}. The experimental values for the atomic volume and the magnetic moment come
from Refs. \onlinecite{endoh71jpsj30:1614,wyckoff63}, and are obtained as extrapolation to room temperature of high temperature data. The values of the bulk modulus are more uncertain and come from Refs. \onlinecite{guillermet89prb40:1521,moruzzi93prb48:7665}.}

  \begin{tabular}{|c|c|c|c|c|c|}
  \hline
   &  \hspace{0.35cm} LDA \hspace{0.35cm}   & \hspace{0.02cm} $U=2.6$ eV \hspace{0.02cm} & \hspace{0.02cm} $U=3.0$ eV \hspace{0.02cm} & \hspace{0.35cm} GGA \hspace{0.35cm} &  \hspace{0.25cm} EXP \hspace{0.25cm} \\
  \hline
  \hline
  \hline
  \begin{tabular}{c|l}
  $ V_0 (\text{a.u.}^3) \: \! $ & \begin{tabular}{l} FP-LMTO \\ [1ex] Ref. \onlinecite{eder00prb61:11492} \end{tabular}
  \end{tabular}
   &
     \begin{tabular}{c}
      69.18 \\ [1ex]
      68.36 \\ 
     \end{tabular}
   &
         \begin{tabular}{c}
          81.17 \\ [1ex]
           \\ 
         \end{tabular}
    &
             \begin{tabular}{c}
              88.61 \\ [1ex]
                \\ 
             \end{tabular}
        &
                 \begin{tabular}{c}
                   \\ [1ex]
                   82.32 \\ 
                 \end{tabular}
         &
         $87.30 \div 87.60$ 
           \\
  \hline
  \hline
  \hline
  \begin{tabular}{c|l}
   $B(\text{GPa})\:\:$ & \begin{tabular}{l} FP-LMTO \\ [1ex] Ref. \onlinecite{eder00prb61:11492} \end{tabular}
  \end{tabular}
  &
     \begin{tabular}{c}
      313 \\ [1ex]
      310 \\ 
     \end{tabular}
   &
         \begin{tabular}{c}
          213 \\ [1ex]
           \\ 
         \end{tabular}
    &
             \begin{tabular}{c}
              88 \\ [1ex]
               \\ 
             \end{tabular}
      &
                 \begin{tabular}{c}
                   \\ [1ex]
                   95 \\ 
                 \end{tabular}
       &
         $90 \div 130$
            \\
  \hline
  \hline
  \hline
  \begin{tabular}{c|l}
   $\mu({\mu_B})\:\:\:\:\:\,$ & \begin{tabular}{l} FP-LMTO \\ [1ex] Ref. \onlinecite{eder00prb61:11492} \end{tabular}
  \end{tabular}
  &
     \begin{tabular}{c}
      0.00 \\ [1ex]
      0.00 \\ 
     \end{tabular}
   &
         \begin{tabular}{c}
          1.74 \\ [1ex]
           \\ 
         \end{tabular}
    &
             \begin{tabular}{c}
              2.30 \\ [1ex]
               \\ 
             \end{tabular}
      &
                 \begin{tabular}{c}
                   \\ [1ex]
                   2.40 \\ 
                 \end{tabular}
       &
          2.30
            \\
   \hline
\end{tabular}
\end{table*}

Consistently with previous calculations, the LDA fails for
$\gamma$-Mn and the differences with the experimental data are
much stronger than for the other transition metals, e.g. Ni
presented above. The atomic volume is underestimated and the bulk
modulus is heavily overestimated. Moreover for $\gamma$-Mn
the change of the exchange-correlation potential from LDA to GGA
does not solve all the problems, and still there is an important
difference between theory and experiments. Does the LDA+DMFT
scheme give a better description? The simulation for the weakest
$U$ seems to underestimate the local Coulomb interaction. The
corrections of equilibrium atomic volume, bulk modulus and magnetic
moment are
good, but they are too small to reproduce the experimental data.
On the other hand the simulation for the strongest $U$ is 
in perfect agreement with the reported values. Nevertheless we
must notice that the quantitative difference of the bulk modulus
between the two LDA+DMFT simulations is surprisingly big. From the
comparison with FP-KKR data, and also looking to the results
for Ni, we see that our value is slightly underestimated because
of the use of the basic DMFT cycle, but we can exclude that this
effect comprehend the whole variation of $B$. We identify this
sensitivity of $B$ to $U$ as another sign of strong correlations.

The reliability of the solver used in the presented
calculations has been checked carefully. In fact the SPTF
solver is a perturbative approach to the Anderson impurity model,
and its application is restricted to systems where the Hubbard $U$
is not bigger than the bandwidth. In this sense $\gamma$-Mn
is a system at the border of the range of applicability, so that
a deep investigation of the behavior of SPTF has been necessary.
Given that the localization of
the $3d$ electrons depends on the atomic volumes, we could expect
that our approximations are not valid for high values of the
lattice constant. We surely exclude this problem since we verified
that this happens only far away from the range of atomic volumes
we were interested in. Another problem we could exclude was the
fact that our approximations could simply collapse for all the
atomic volumes driven by the strength of $U$. In fact we have
studied intermediate values of $U$ between $U=2.6 \text{ eV}$
and $U=3.0 \text{ eV}$ and all the physical properties have
shown a regular behavior, including the bulk modulus $B$.

While we focused our analysis mainly on the anti-ferromagnetic
phase, we tried to get more insight into the role of magnetism
in $\gamma$-Mn through LDA+DMFT simulations of the
non-magnetic phase. The results are quite interesting: the energy
versus lattice constant curve (not shown here) has a regular
parabolic shape with an equilibrium atomic volume 
$V_0=85.91 \text{ a.u.}^3$, intermediate to the equilibrium atomic
volume of the LDA+DMFT simulation for the antiferromagnetic
phase. Obviously this is a consequence of the constrained zero magnetic
moment, and no quenching of the Galitskii-Migdal energy can
appear. The increasing strength of the correlation energy
is observable also in a huge drop of the bulk modulus
with respect to its bare LDA value: $B = 57 \text{ GPa}$, perfectly
consistent with the already mentioned experimental data for
$\gamma$-MnCu alloys\cite{tsunoda84jpsj53:359}, after extrapolation
to zero content of Cu at room temperature.
As before we have checked whether the SPTF solver is applicable
or not to our system. We have found that our approximations lose
validity for atomic volumes larger than $100 \text{ a.u.}^3$: the
localization effects are heavily overestimated and the crystal
tends to collapse into an atomistic system. Fortunately this threshold
is well above the equilibrium values, so that we can
still consider our results as reliable.

\section{Conclusions}
In this paper we have presented two different total energy implementations for the LDA+DMFT method, using the SPTF solver for the solution of the local problem. Our codes  have  been tested through the study of the ground-state properties of fcc Ni. The results have been very encouraging, showing good agreement with experimental data and in a particular a weak dependence on the implementation or on the choice of the local orbitals. Furthermore a tendency of the basic LDA+DMFT cycle to underestimate
the bulk modulus with respect to the fully self-consistent cycle has been observed.

The main scientific aim of this paper has been the analysis of the role of local correlations in $\gamma$-Mn. Clear signs of strong correlations have been found and the LDA+DMFT method has been shown the ability to treat the non-magnetic and anti-ferromagnetic phases on the same footing, improving considerably the results obtained with usual one-particle approximations.

Finally the results presented here stimulate future research. The main question concerns the origin of a kink in the total energy curves and the role of the tetragonal distortion of the fcc lattice on the correlation effects of the antiferromagnetic phase of $\gamma$-Mn. This last study can be particularly interesting for the calculation of the elastic properties.
In addition the influence of the choice of the solver on the description of $\gamma$-Mn needs more investigation.

\section*{ACKNOWLEDGMENTS}
This work was sponsored by the Stichting Nationale Computerfaciliteiten (National Computing Facilities Foundation, NCF) for the use of the supercomputer facilities, with financial support from the Nederlandse Organisatie voor Wetenschappelijk Onderzoek (Netherlands Organization for Scientific Research, NWO). Furthermore the programming part of this work was carried out
under the HPC-EUROPA++ project (application number: 1122), with the support of the European Community - Research Infrastructure Action of the FP7. Fundamental support was also given by the Deutsche Forschungsgemeinschaft within
the priority program ``Moderne und universelle first-principles-Methoden
f\"ur Mehrelektronensysteme in Chemie und Physik'' (SPP 1145/2).


\end{document}